\documentclass[lettersize,journal]{IEEEtran}
\usepackage{amsmath,amsfonts}
\usepackage{algorithmic}
\usepackage{algorithm}
\usepackage{array}
\usepackage[font=footnotesize, labelfont=bf, justification=raggedright, singlelinecheck=false]{caption}
\usepackage{subcaption}
\usepackage{textcomp}
\usepackage{stfloats}
\usepackage{url}
\usepackage{multirow}
\usepackage{makecell}
\usepackage{diagbox}
\usepackage{verbatim}
\usepackage{graphicx}
\usepackage{cite}
\usepackage{xcolor}
\newcommand{\revise}{\textcolor{black}}
\begin{document}

\title{Integrating Sensing into Covert Communications: Opportunities and Challenges}


  \author{
    \IEEEauthorblockN{Jun~Wu,  Haoyuan~Pan,~\IEEEmembership{Member,~IEEE}, 
    Gaosheng Zhao, \\
    Dong~In~Kim,~\IEEEmembership{Life~Fellow,~IEEE}, and~Tse-Tin~Chan,~\IEEEmembership{Member,~IEEE}
  }
  \thanks{This work was supported in part by the National Natural Science Foundation of China under Grant 62371302 and in part by the Research Matching Grant Scheme from the Research Grants Council of Hong Kong. This work of D. I. Kim was supported by the National Research Foundation of Korea (NRF) grant funded by the Korean Government (MSIT) under Grants RS-2026-25608310 and RS-2026-25470140. \textit{(Corresponding author: Tse-Tin Chan.)}

  J. Wu and T.-T. Chan are with the Department of Mathematics and Information Technology, The Education University of Hong Kong, Hong Kong, China (e-mail: wjun@eduhk.hk; tsetinchan@eduhk.hk). 
  

H. Pan is with the College of Computer Science and Software Engineering, Shenzhen University, Shenzhen, China (e-mail: hypan@szu.edu.cn).

G. Zhao and D. I. Kim are with the Department of Electrical and Computer Engineering, Sungkyunkwan University, Suwon 16419, South Korea (e-mail: \{gaosheng, dongin\}@skku.edu).

  }
  }
\maketitle

\begin{abstract}
Covert communications aim to hide the existence of wireless transmissions from unauthorized adversaries. However, conventional designs based on blind interference or passive uncertainty can be ineffective in dynamic propagation environments. This article investigates sensing-empowered covert communications, where adversary and environmental information are used to guide transmission and jamming control. We show how sensing changes covert system design from passive concealment to state-aware decision-making, while also introducing new challenges related to exposure and resource consumption. We further discuss several intelligent sensing paradigms that extract task-relevant information with limited active probing. A case study in low-altitude wireless networks illustrates that sensing-assisted beamforming can improve task-relevant spatial resource utilization under communication-rate and leakage-power constraints. Finally, several open issues are discussed to support more adaptive covert wireless systems. 
\end{abstract}



\section{Introduction}

Covert communications have attracted growing interest as an important physical-layer technique for low-probability-of-detection wireless services\cite{chen2023covert,zhao11359112}. Unlike conventional physical-layer security, which primarily aims to prevent unauthorized decoding, covert communication aims to hide the physical existence of the transmission. This requirement is important for security-sensitive wireless applications, where even exposure of communication activity may pose a critical operational risk\cite{su2025integrating}. In such systems, the unauthorized monitor performs statistical detection based on the received signal. In contrast, the legitimate system must ensure that the presence of covert data transmission is statistically difficult to distinguish from its absence.

The fundamental difficulty of covert communication comes from the coupling between reliable transmission and low detectability. In particular, a stronger transmitted signal can improve the legitimate link, but it also increases the possibility of being observed by the monitor. \revise{To address this issue, existing covert designs typically exploit uncertainty at the monitor to reduce the statistical difference between observations under transmission and silence. Representative approaches include noise and environmental interference uncertainty \cite{he2017covert} and artificial-noise (AN)-assisted transmission \cite{sobers2017covert}. More recently, sensing-aided covert communications have exploited adversary and environmental information to guide transmission and interference control \cite{jun2025aerial}. These studies demonstrate the potential of sensing for improving adversary awareness, but mainly focus on specific sensing-assisted transmission or jamming mechanisms rather than a systematic sensing-empowered framework.}

However, the effectiveness of covert transmission strongly depends on the accuracy and freshness of the available adversary information. Without reliable knowledge of the warden channel, friendly jamming may lack spatial guidance, consuming excessive power and impairing the legitimate receiver while providing only marginal improvement in covertness. This issue is further amplified in dynamic wireless environments, where mobility and blockage can quickly reshape the detection link. \revise{As a result, a fixed covert policy derived from long-term statistics may fail to capture the actual exposure risk. Sensing provides a practical means of addressing this limitation. By acquiring information about the monitor and the surrounding propagation environment, the legitimate system can adapt its transmission policy to the actual monitoring risk. For example, the legitimate transmitter may reduce leakage toward high-risk directions or exploit favorable propagation conditions for covert data delivery. In this sense, sensing enables a shift from blind uncertainty exploitation to state-aware covert design.}

Nevertheless, we note that sensing cannot be directly incorporated into covert communication without careful design\cite{wang2024sensing}. On the one hand, active sensing requires signal emission, which may inadvertently increase the exposure of the legitimate transmitter.  On the other hand, sensing functions also consume time and power resources that could otherwise be used for data transmission. Moreover, sensing information may become outdated in mobile scenarios due to processing delays and channel variations. Therefore, sensing-empowered covert communication introduces a new design problem, where sensing accuracy, communication reliability, and covertness must be jointly considered.

Motivated by the above, this article provides a systematic study of sensing-empowered covert communications. The main contributions of this article are summarized as follows:

\begin{itemize}
    \item We present representative architectures for sensing-empowered covert communications. These architectures show how sensing can support adversary-aware transmission and friendly jamming control. We identify the new design considerations introduced by sensing. In particular, we discuss how sensing changes the performance metrics, enlarges the optimization degrees of freedom (DoFs), and creates coupled constraints between sensing reliability and covertness.

    \item We discuss intelligent sensing paradigms for covert communications. Instead of pursuing maximum sensing accuracy, these paradigms focus on extracting decision-relevant information under limited probing and exposure budgets.

    \item We present a case study of low-altitude wireless networks (LAWNs) to illustrate the benefits of intelligent sensing-assisted covert transmission. Finally, the promising directions toward sensing-assisted covert communications are outlined.
\end{itemize}

\section{Sensing-Empowered Covert Communications}
\label{sec:sensing_empowered_covert}

\begin{figure*}[t]
    \centering
    \begin{subfigure}[t]{0.32\textwidth}
        \centering
        \includegraphics[width=\linewidth]{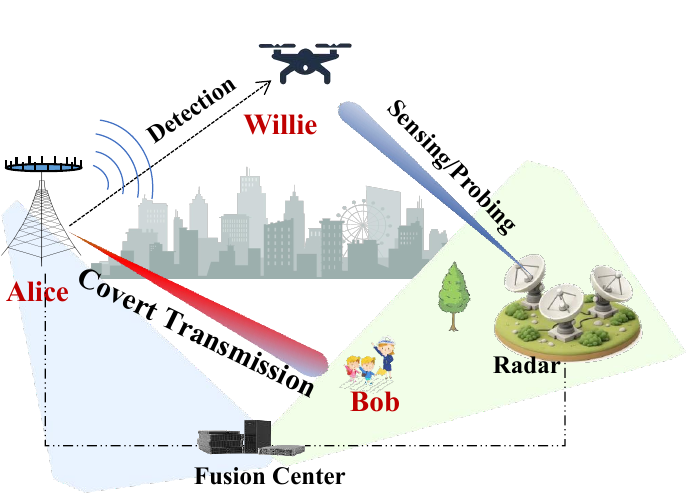}
        \caption{Sensing-then-covert transmission.}
        \label{fig:sensing_then_transmission}
    \end{subfigure}
    \hfill
    \begin{subfigure}[t]{0.32\textwidth}
        \centering
        \includegraphics[width=\linewidth]{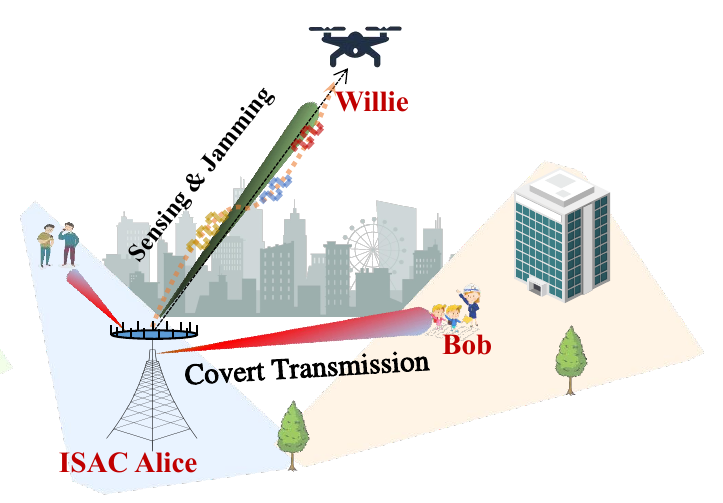}
        \caption{Integrated sensing and jamming.}
        \label{fig:integrated_sensing_jamming}
    \end{subfigure}
    \hfill
    \begin{subfigure}[t]{0.32\textwidth}
        \centering
        \includegraphics[width=\linewidth]{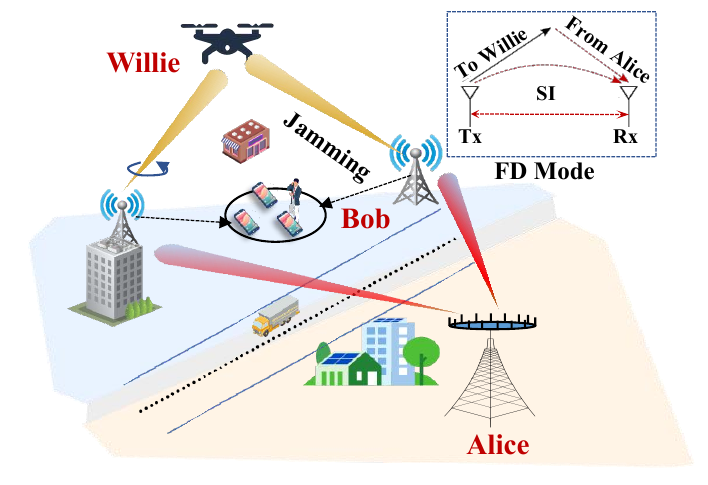}
        \caption{Full-duplex receiver-assisted covert transmission.}
        \label{fig:receiver_assisted_fd_covert}
    \end{subfigure}
    \caption{Representative modes of sensing-empowered covert communications.}
    \label{fig:sensing_empowered_covert_modes}
\end{figure*}



    Covert communications aim to prevent a potential monitor, Willie, from determining whether a legitimate transmitter, say Alice, is transmitting covert information to a legitimate receiver, Bob. In sensing-empowered covert communications, the hypotheses should be defined with respect to the presence or absence of Alice's covert data signal, rather than the presence or absence of all auxiliary sensing or jamming emissions. Specifically, Willie formulates the monitoring process as a binary hypothesis testing problem, i.e.,
\begin{itemize}
    \item \textbf{Null hypothesis $\mathcal{H}_{0}$}: Alice does not transmit covert information to Bob. In this case, Willie's observation may consist of background noise, environmental uncertainty, and auxiliary sensing or jamming waveforms that are not directly correlated with Alice's covert-data activity.

    \item \textbf{Alternative hypothesis $\mathcal{H}_{1}$}: Alice transmits covert information to Bob. In this case, Willie's observation contains not only the background and auxiliary components, but also the signal leakage induced by Alice's covert data transmission.
\end{itemize}
In the considered architectures, sensing is exploited in two complementary ways. First, sensing provides Willie-related side information, such as Willie's location/direction, blockage state, and coarse indicators of the Alice--Willie monitoring channel, which supports state-aware transmission, beamforming, and jamming control. Second, a non-information-bearing sensing or probing waveform can be reused as a cover signal or friendly interference to mask the leakage of Alice's covert signal at Willie. We do not assume an independent sensing-only transmission from Alice before covert data delivery. Instead, sensing is performed either by external sensing nodes, by the full-duplex receiver Bob, or through Alice's joint sensing/communication waveform in a manner decoupled from the covert-data state. Therefore, the auxiliary sensing/jamming waveform should be designed so that its observable statistics and scheduling do not provide a direct indicator of Alice's covert data activity. Consequently, the legitimate system should jointly design the covert-data transmission and auxiliary sensing/jamming waveforms such that Willie's observations under $\mathcal{H}_{0}$ and $\mathcal{H}_{1}$ are statistically difficult to distinguish~\cite{xu10938035,kimhieu2023joint}.

To address these challenges, sensing can be exploited to provide Alice or the legitimate system with Willie-related information. Subsequently, the system can steer interference toward Willie and adjust the covert beam based on the actual detection risk. In this way, sensing transforms blind interference injection into targeted interference control. As shown in Fig.~\ref{fig:sensing_empowered_covert_modes}, sensing-aided covert communications can be realized through the following representative modes, including sensing-then-transmission, integrated sensing and jamming, and receiver-assisted full-duplex covert transmission.

\subsection{Sensing-Then-Covert Transmission}
\label{subsec:sensing_then_transmission}

The sensing-then-transmission paradigm is illustrated in Fig.~\ref{fig:sensing_then_transmission}. In this mode, sensing and covert transmission are performed sequentially. The legitimate system first employs external sensing nodes to characterize the surrounding radio environment. The sensing results are collected and processed by a fusion center, which extracts useful information such as Willie's location\cite{wang2024sensing}. Based on this information, Alice adapts her covert transmission policy toward Bob.

This paradigm is particularly useful when dedicated sensing infrastructure is available. Specifically, the system can exploit high-resolution sensing devices to obtain a more accurate estimate of the adversary state. For example, if sensing indicates that Willie is located close to Alice, Alice may need to adopt a worst-case strategy to guarantee covertness, or even temporarily suspend transmission. By contrast, if Willie is blocked by obstacles or immersed in strong environmental interference, Alice may achieve a higher covert rate while still satisfying the covertness constraint. Therefore, sensing enables the transformation of passive covert communication into active adversary-aware transmission designs. Instead of relying only on statistical assumptions about Willie, the system actively estimates the adversary state, enabling more flexible covert beamforming, power control, and time-frequency resource allocation. 

Despite these advantages, this paradigm also suffers from practical limitations. On the one hand, the sensing phase consumes computational resources before data delivery, thereby inevitably increasing system delay and reducing the available resource budget for covert transmission. When the monitor Willie is mobile with varying channels, the transmission policy derived from the sensed state may no longer match the actual detection condition at Willie. As a consequence, the sequential design should carefully balance sensing overhead and information freshness against the resulting covertness gain.

\subsection{Integrated Sensing and Jamming for Covert Transmission}
\label{subsec:sensing_jamming_integrated}

To tackle the above challenges, a promising trend is to develop integrated sensing and communication (ISAC)-based covert transmission, as illustrated in Fig.~\ref{fig:integrated_sensing_jamming}. Unlike the sensing-then-transmission paradigm, ISAC does not require a separate sensing phase before data delivery. Instead, sensing and communication are supported by a unified radio platform with shared waveforms, spectrum, and hardware resources\cite{jun2025aerial}. By doing so, Alice can update Willie-related information with lower system latency.


    More importantly, sensing under this framework is no longer only a preliminary operation for obtaining environmental information. Since the sensing waveform may be emitted during the covert communication process, it also affects the signal Willie receives. Nevertheless, the sensing waveform does not automatically improve covertness. If its structure is known to Willie or if its observable statistics and scheduling are correlated with Alice's covert-data activity, Willie may exploit it as an additional detection feature. Therefore, the sensing waveform should be deliberately designed as a cover signal. By randomizing its waveform statistics, decoupling its observable scheduling from the covert-data state, and steering its energy toward Willie while limiting its impact on Bob, the legitimate system can increase Willie's observation uncertainty and mask Alice's covert signal. In this sense, Alice can be regarded as a triple-functional transmitter that communicates with Bob, senses Willie-related states, and uses the designed sensing waveform to reduce the statistical distinguishability between $\mathcal{H}_{0}$ and $\mathcal{H}_{1}$ at Willie. This architecture offers two potential advantages under proper waveform design. First, the reuse of sensing waveforms can reduce the need for dedicated AN, thereby improving energy and spectrum efficiency. Second, the acquired sensing information enables the cover signal or friendly interference to be spatially focused toward Willie rather than blindly spread over the network.

Nevertheless, this mode creates a tight coupling between sensing and covert communication. Increasing the sensing power can improve adversary-state estimation and strengthen the interference observed at Willie, but it may also raise Alice's exposure risk. Reducing the sensing power, on the other hand, may alleviate emission leakage but can degrade both sensing accuracy and covertness enhancement.

\subsection{Full-Duplex Receiver-Assisted  Covert Transmission}

Departing from conventional transmitter-centric paradigms, this architecture structurally shifts the environmental shaping burden to the full-duplex ISAC receiver Bob. In this configuration, Bob proactively dictates the radio environment by broadcasting a dual-purpose probing-jamming waveform, enabling Alice to maintain a low-power transmission\cite{shahzad2018achieving}.

Crucially, this architecture exploits the inherent vulnerability asymmetry prevalent in covert scenarios. By shifting the dominant probing/jamming emissions to the receiver, the system can partially decouple the physical detection risk from the vulnerable transmitter, thereby reducing Alice's direct exposure.  Bob's emitted waveform plays a dual role in this design. It supports adversary sensing while simultaneously serving as AN from Willie's perspective. Since Bob has prior knowledge of this waveform, Bob can suppress the resulting self-interference (SI) and recover Alice's covert message. In contrast, Willie observes a superposition of these signals, which increases the detection uncertainty between $\mathcal{H}_{0}$ and $\mathcal{H}_{1}$. \revise{However, the receiver-driven framework remains constrained by residual SI after self-interference cancellation (SIC), which degrades Bob's effective signal-to-interference-plus-noise ratio (SINR). As the residual SI increases, Alice may require higher transmit power to maintain reliable communication, thereby increasing signal leakage toward Willie and making the covertness constraint more difficult to satisfy.} 




\subsection{New Insights for Sensing-Aided Covert Communications}
\label{subsec:new_insights_sensing_aided}

In this subsection, we highlight the new insights enabled by sensing-aided covert communications. Specifically, the integration of active sensing fundamentally changes system design in terms of performance metrics, optimization degrees of freedom, constraints, and system tradeoffs.

First, sensing introduces composite performance metrics. Traditional covert communications are often evaluated through detection-theoretic metrics, such as Willie's false-alarm and miss-detection probabilities, or through information-theoretic limits, such as the square-root law in large-blocklength regimes. In sensing-integrated systems, performance evaluation should further account for sensing accuracy, such as the Cram\'er-Rao lower bound (CRLB). Therefore, the design should account for sensing quality in addition to communication performance and covertness. Moreover, active sensing expands the optimization DoFs. Conventional covert designs mainly rely on transmit power control, beamforming design, or AN injection. By contrast, sensing-aided architectures provide additional spatial and temporal DoF through sensing covariance optimization and duration allocation.  However, sensing-aided designs are subject to coupled constraints because the system must jointly satisfy requirements for sensing resources, communication reliability, and covertness.

 In addition, sensing introduces a fundamental tradeoff between sensing and covertness. More accurate sensing can improve adversary awareness and support more efficient covert adaptation, but it also consumes resources that could otherwise be used for data transmission or friendly jamming. Conversely, allocating more resources to covert data transmission may increase the achievable rate, but it can also enlarge the statistical difference between the silent and active hypotheses at Willie. Therefore, sensing-aided covert communication should jointly balance sensing accuracy and covert performance.

\section{Intelligent Sensing Paradigms for Covert Communications}
\label{sec:intelligent_sensing_paradigm}

\begin{figure*}[t]
  \centering
    \includegraphics[width=\linewidth]{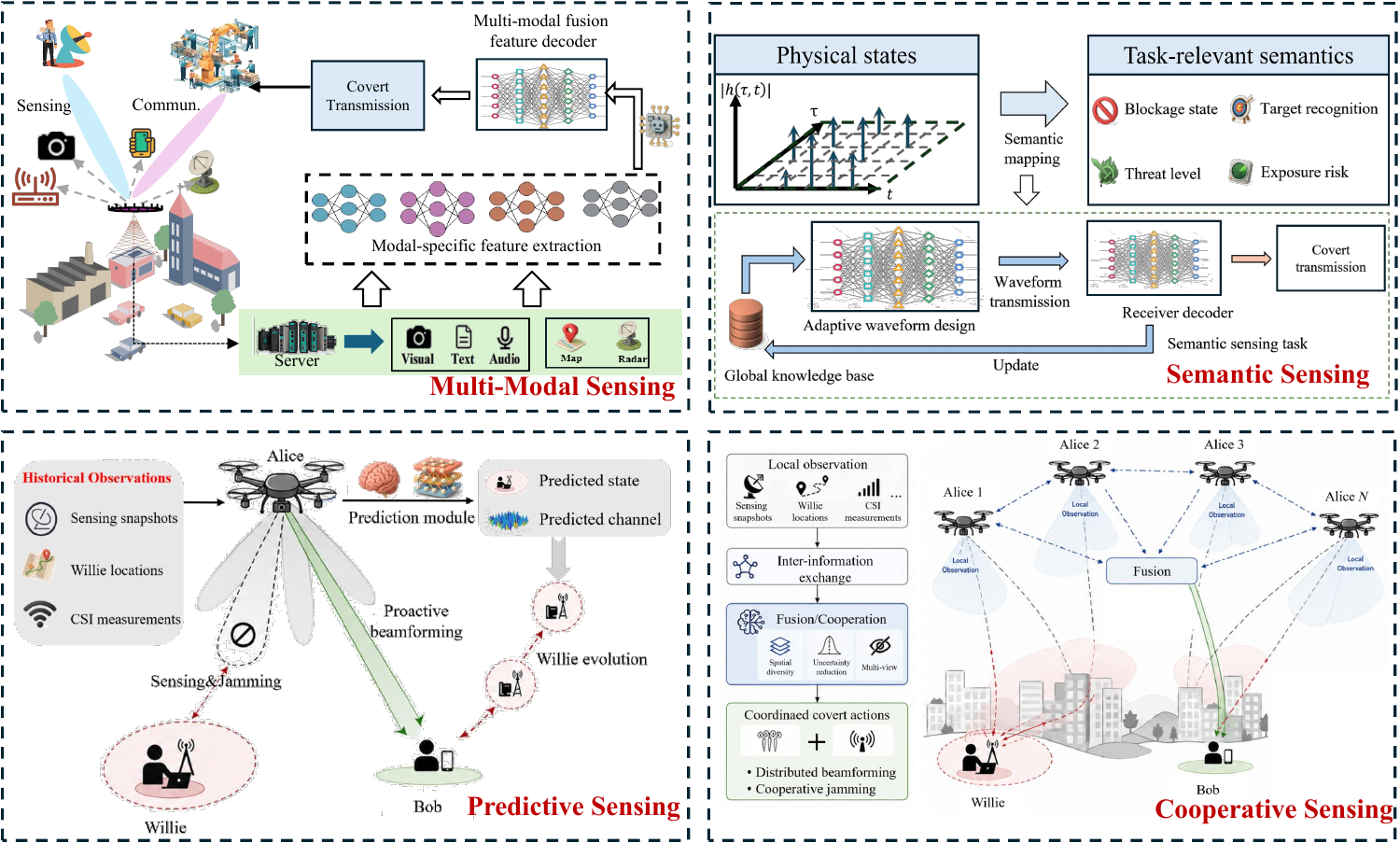}
\caption{Intelligent sensing paradigms for covert communications. Predictive sensing anticipates adversary mobility to support proactive beamforming, while semantic sensing extracts task-relevant features to reduce unnecessary probing. Cooperative sensing exploits spatial diversity to mitigate sensing blind spots, while multi-modal sensing fuses cross-domain information to enable robust and low-exposure situational awareness.}
  \label{fig:intelligent_sensing}
\end{figure*}

\revise{The integration of active sensing capabilities transforms blind interference into a controllable resource for covert communications. However, conventional ISAC systems often rely on high-power, wide-bandwidth transmissions to enable accurate multi-dimensional parameter estimation, thereby increasing the likelihood that the legitimate transmitter is detected by Willie. Therefore, future covert networks require a fundamental shift from conventional physical measurements to an intelligent sensing paradigm \cite{li2025intelligent}. The key objective is not to maximize sensing accuracy alone, but to acquire the most useful information for covert decision-making under strict covert constraints. In this context, multi-modal and cooperative sensing can improve Willie awareness while reducing the limitations of individual observations, whereas semantic and predictive sensing can reduce unnecessary probing and mitigate information aging. These paradigms are therefore adapted to covert communications to improve adversary awareness and transmission decisions without introducing excessive sensing-induced exposure, as illustrated in Fig.~\ref{fig:intelligent_sensing}.}

\subsection{Multi-Modal Sensing}

\label{subsec:multi_modal_sensing}

Multi-modal sensing provides complementary environmental information for communication design~\cite{cheng2023intelligent}. Building on this capability, we envision its use to mitigate the sensing-exposure dilemma in covert communications. In traditional radio-based sensing, accurate environmental perception often relies on frequent radio-frequency probing, which may increase the probability of exposing Alice. By incorporating heterogeneous information sources, such as visual observations, environmental maps, and historical mobility data, the legitimate system can infer the adversary's state and propagation conditions with reduced reliance on active radio emissions. Another important benefit of multi-modal sensing is improved robustness. Covert transmission decisions depend on the knowledge of Willie's location and monitoring capability. However, a single radar sensing modality may be affected by blockage and non-line-of-sight (NLoS) propagation. Multi-modal fusion can mitigate these limitations by exploiting complementary information from different sources. For example, visual observations can assist in tracking Willie's movement, while wireless measurements can capture variations in link quality. Environmental maps and historical observations can further help predict blockage conditions and monitoring risks. Therefore, multi-modal sensing can support the construction of a more reliable exposure-risk map.

 Multi-modal sensing can further reveal the mapping relationship between physical-domain observations and covert communication performance. By jointly processing the multi-modal observations, the legitimate system can learn how physical features affect the Alice-Willie channel and Willie's monitoring capability. This allows Alice to infer covertly relevant information with reduced reliance on active radio probing, thereby supporting beamforming and power control at a lower exposure cost. 

\subsection{Semantic Sensing} \label{subsec:semantic_sensing}
While multi-modal integration provides a comprehensive representation of the physical environment, the resulting high-dimensional data may introduce substantial processing and signaling overhead. To bridge the gap between raw measurements and actionable transmission strategies, semantic sensing shifts the perception objective from reconstruction-oriented accuracy to task-oriented interpretation\cite{zhang2026semantic}. Specifically, conventional sensing architectures often focus on estimating physical parameters, such as delays, Doppler shifts, and angles of arrival. However, these parameters may contain redundant information that is not directly relevant to covert transmission. Instead of performing full channel reconstruction, semantic sensing extracts compact covertness-relevant variables. For instance, rather than estimating the exact coordinates of all scatterers, the legitimate system may infer task variables, such as the risk indicator and the threat level.

To realize this abstraction, semantic sensing requires dedicated transceiver designs. At the sensing transmitter, the waveform can be optimized to emphasize task-discriminative environmental features while suppressing unnecessary probing. Meanwhile, a semantic decoder at the receiver maps the received physical-layer observations directly into covertness-relevant semantic variables, avoiding explicit high-dimensional channel reconstruction. \revise{The implementation complexity of the semantic decoder depends primarily on the dimensionality of the sensing observations and the adopted inference architecture. Lightweight feature extraction and task-oriented inference can reduce the processing burden.} This design reduces both signaling overhead and sensing-induced exposure, which is particularly useful in resource-constrained covert networks. Nevertheless, the semantic abstraction should be carefully controlled, since excessive compression may discard critical exposure cues, while insufficient compression may weaken the benefits of semantic sensing.

\subsection{Predictive Sensing}
\label{subsec:predictive_sensing}


While semantic sensing alleviates the data processing overhead, highly mobile environments further require delay-sensitive covert adaptation. In dynamic scenarios such as low-altitude wireless networks, the network topology and propagation channels may vary rapidly\cite{wu11503149}. If covert resource allocation is designed only based on past sensing observations, the estimate of Willie's state may become outdated, leading to unintended signal leakage. Toward this end, predictive sensing shifts covert communications from reactive decision-making to proactive risk management over time. Instead of relying solely on instantaneous state estimation, the system predicts the future evolution of Willie's mobility, wireless channels, and exposure risks.  

The core mechanism of predictive sensing relies on exploiting historical sensing observations to anticipate future transmission conditions. By extracting temporal correlations from historical channel state information (CSI) and past adversary locations, the legitimate system can forecast the non-stationary evolution of the wireless environment. This enables the transmitter to anticipate transient vulnerabilities and proactively optimize its predictive beamforming and resource allocation before imminent threats. For example, Alice can utilize predicted low-risk time windows, e.g., deep fading at Willie, to enhance covert data delivery. Conversely, it may reduce transmit power or postpone transmission when a high detection risk is anticipated.
However, the effectiveness of predictive sensing is strictly constrained by prediction uncertainty. Because Willie's exact behavior and the dynamic radio environment cannot be perfectly predicted, estimation errors may lead to inappropriate transmission decisions and severe degradation of covertness. 
In high-mobility LAWNs, predictive beamforming may also suffer from short channel coherence time and frequent state-update overhead. A compact delay-Doppler-domain representation can provide more structured mobility-related features for prediction than directly tracking rapidly varying instantaneous channel coefficients\cite{zhang2024predictive}.

\subsection{Cooperative Sensing}
\label{subsec:cooperative_sensing}

Although the aforementioned paradigms improve the sensing capability of a legitimate node, they remain limited by the node's local field of view. In complicated propagation environments, physical blockages may create sensing blind spots and reduce the available environmental information. If Willie is located in an occluded region, single-node sensing may fail to obtain sufficient adversary information, degrading the reliability of covert decision-making.

Cooperative sensing addresses the limitations of individual sensing nodes by forming a distributed perception network\cite{jun2025aerial}. Through reliable communication links, multiple legitimate nodes can share local observations to obtain broader situational awareness. Depending on the resource constraints, information exchange can be performed at the raw-data, feature, or semantic level. In particular, sharing compressed feature maps extracted by learning models can provide useful spatial information while reducing the communication burden on the legitimate network. By fusing observations from multiple viewpoints, the legitimate network can mitigate sensing blind spots and enhance Willie detection in NLoS or occluded regions, thereby fully exploiting cross-node spatial diversity. Such spatial awareness further supports distributed covert countermeasures. For example, legitimate nodes can coordinate their beamforming directions to suppress signal leakage toward Willie, or jointly design friendly jamming to degrade the monitoring link. 

\section{Case Study: Intelligent Sensing-Assisted Covert Transmission in a LAWN}

\subsection{System Description}

\revise{Building on the sensing paradigms in Section III, this case study illustrates predictive and semantic beamforming in a low-altitude wireless network.} \revise{We consider a simplified line-of-sight case study with Alice at an altitude of 50 m, equipped with a 32-element uniform linear array. Bob is located at $-30^\circ$, while three Willie-related directions at $43^\circ$, $44^\circ$, and $45^\circ$ are assigned different sensing priorities.} Due to sensing and processing delays, the currently sensed Willie direction may become outdated during transmission, which motivates the use of predictive beamforming to compensate for mobility-induced mismatch. Meanwhile, covertness-oriented design does not require a full reconstruction of the physical environment. Instead, the key is to identify covertness-relevant propagation features, such as leakage-dominant directions and blockage-assisted low-risk regions. In this case study, semantic sensing is implemented through semantic beamforming, where the transmitter maps sensing observations into covertness-relevant angular regions and then allocates sensing and transmission resources based on their relevance to covert decision-making. Compared with full-region sensing, this design reduces unnecessary exposure while preserving spatial DoFs for Bob's communication link.

\revise{Three schemes are compared to illustrate how intelligent sensing balances task-relevant illumination and communication performance under covertness constraints in LAWNs. The central Willie-related direction is treated as the given predicted direction in this case study. Conventional sensing uses a fixed beam toward this direction, while predictive beamforming maximizes sensing power toward it. Semantic beamforming maximizes task-weighted sensing power across all three Willie-related directions.} 

\subsection{Results and Discussion}

\begin{figure}[htbp]
    \centering
    \begin{subfigure}[b]{\linewidth}
        \centering
        \includegraphics[width=\linewidth]{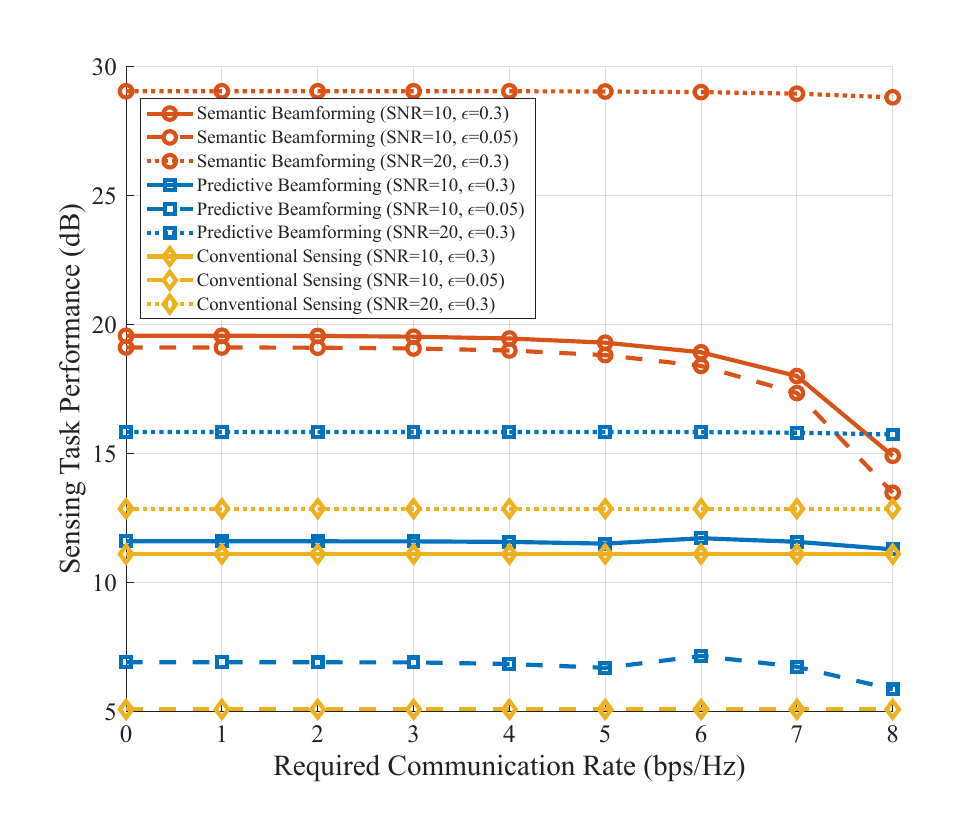}
        \caption{Sensing task performance under communication-rate constraints.}
        \label{fig:sensing_vs_comm_rate}
    \end{subfigure}
    
    \begin{subfigure}[b]{\linewidth}
        \centering
        \includegraphics[width=\linewidth]{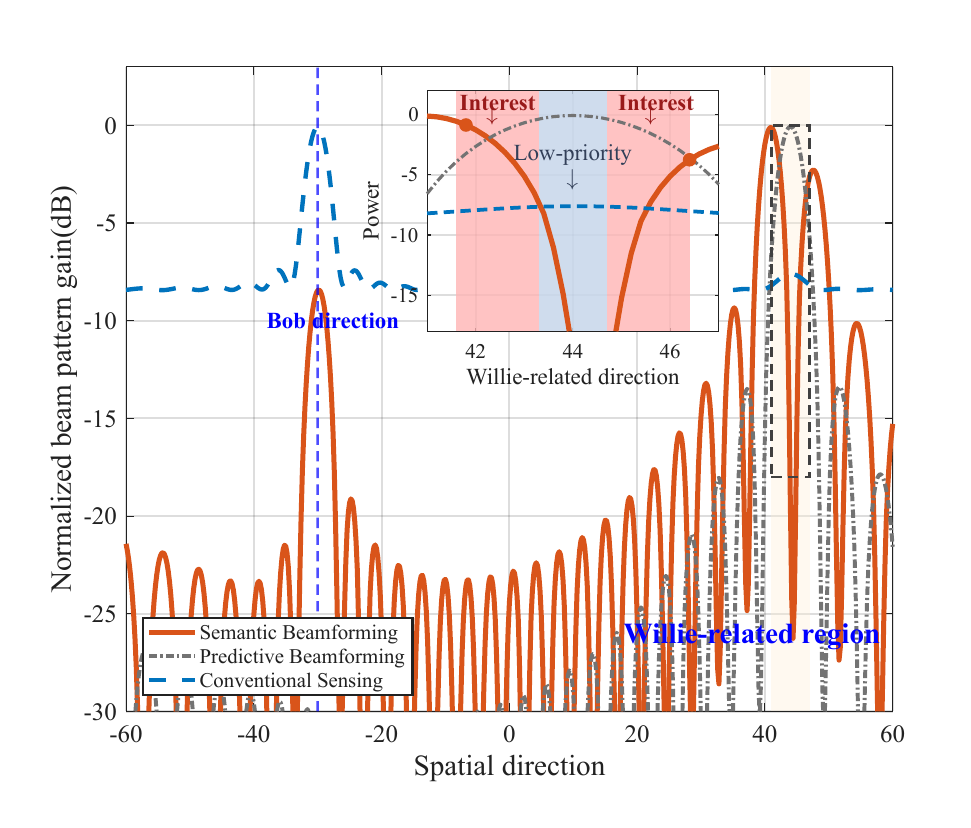}
        \caption{Spatial beampatterns of the compared sensing strategies.}
        \label{fig:spatial_beam_pattern}
    \end{subfigure}
    
        \caption{Intelligent sensing for covert communications in urban LAWNs.}
    \label{fig:simulation_results}
\end{figure}

\revise{Fig. \ref{fig:simulation_results}(a) compares task-weighted illumination under communication-rate constraints, while Fig.~\ref{fig:simulation_results}(b) shows the spatial beam patterns of the three sensing strategies.} 
\revise{The sensing task performance in Fig.~\ref{fig:simulation_results}(a) is the sum of the transmit beampattern powers in the three Willie-related directions, weighted by their predefined sensing priorities. All three schemes are evaluated using the same directions and weights, with higher values indicating stronger task-relevant illumination.} The signal-to-noise ratio (SNR) denotes the transmit SNR budget at Alice, defined as the maximum transmit power normalized by the noise variance, while $\epsilon$ represents the covertness tolerance mapped to a leakage-power constraint at the Willie-related direction. \revise{For predictive and semantic beamforming, moderate communication-rate requirements leave more power for sensing, whereas higher requirements shift power toward Bob, reducing task-relevant illumination under the fixed transmit-power budget and leakage constraint toward Willie. Semantic beamforming improves sensing effectiveness by prioritizing covertness-relevant directions in the spatial power allocation.}\par


\revise{The inset in Fig.~\ref{fig:simulation_results}(b) shows that semantic beamforming provides stronger illumination in the two higher-priority side regions than in the low-priority central region. Compared with conventional sensing, predictive beamforming optimizes sensing power toward the given predicted direction, while semantic beamforming focuses on task-relevant information to support covert decision-making. The practical implication is to use directional information and task priorities to guide spatial resource allocation. In deployment, predictive beamforming relies on accurate direction information, while semantic beamforming relies on sensing priorities that reflect the current task.}
\revise{Nevertheless, sensing errors and imperfect CSI may cause Willie-state mismatch and beam misalignment, thereby increasing unintended signal leakage. Meanwhile, predictive and semantic sensing introduce additional computational and energy costs for state prediction, feature extraction, and inference, while cooperative sensing may incur increasing signaling and coordination overhead in dense networks. Moreover, a sophisticated Willie with additional side information may exploit the auxiliary sensing or jamming waveform as an additional detection feature.}

\section{Future Directions}
\label{sec:future_directions}

\subsection{AI-Enabled Covert Intelligence}
\label{subsec:ai_enabled_covert_intelligence}

Artificial intelligence (AI) is expected to become a key enabler for sensing-empowered covert communications, especially in highly dynamic and partially observable wireless environments. By learning from heterogeneous sensing observations, AI can help infer adversary behavior, predict exposure risk, estimate channel evolution, and identify favorable opportunities for covert transmission. Instead of relying only on static detection and optimization models, AI-enabled methods can support adaptive beamforming, jamming control, sensing scheduling, trajectory planning, and covertness-aware resource allocation. However, AI must be carefully integrated with detection-theoretic constraints, since unreliable inference or unsafe exploration may expose the legitimate transmitter. 

\subsection{Spectrum Map-Aided Covert Networking}
\label{subsec:spectrum_map_aided_covert_networking}

 By capturing the distribution of spectrum occupancy and channel quality over space and time, spectrum maps provide a structured view of the wireless environment. When enriched with blockage information and monitoring risk, spectrum maps can help the legitimate transmitter identify low-risk transmission regions and exploit ambient masking opportunities. They can also support the selection of sensing nodes and the coordination of beamforming or jamming strategies. However, spectrum-map construction requires continuous sensing and information exchange, which may introduce additional overhead and exposure risk. 
These requirements motivate the development of lightweight and uncertainty-aware spectrum-map construction frameworks tailored to covert communication scenarios. Moreover, the confidentiality of spectrum maps needs to be explicitly protected, since compromised map information may reveal legitimate transmission strategies or distort covert decision-making.
\vspace{-0.3cm}
\revise{
\subsection{Cell-Free Covert Communications
}
Cell-free networks provide a promising architecture for sensing-empowered covert communications by enabling distributed access points to jointly serve legitimate users without conventional cell boundaries. Their spatially distributed observations can improve Willie-state awareness, while cooperative beamforming and jamming can suppress signal leakage toward Willie without relying on a single transmitter. However, such gains require timely sensing-information exchange and coordination among access points, which introduce additional signaling overhead and potential exposure. In this context, lightweight cooperation and covertness-aware distributed processing are essential for realizing scalable cell-free covert networks.}


\vspace{-0.5cm}
\section{Conclusions}
\label{sec:conclusions}

    This article investigated sensing-empowered covert communications, where environmental and adversary-related sensing information can support state-aware transmission and jamming control. We discussed how sensing shifts covert design from blind exploitation of uncertainty toward adaptive decision-making, while introducing new challenges in exposure, resource allocation, and information freshness. We further examined intelligent sensing paradigms that acquire decision-relevant information under limited probing overhead. A LAWN case study illustrated the potential of sensing-assisted beamforming to improve task-relevant spatial resource utilization under communication-rate and leakage-power constraints. Finally, several future research directions were outlined to support more adaptive and robust covert wireless systems.

\bibliographystyle{ieeetr}
\bibliography{reference}
\end{document}